# Unsupervised Anomaly Detection in MR Images using Multi-Contrast Information

Byungjai Kim[1], Kinam Kwon[2], Changheun Oh[1] and Hyunwook Park[1*]

[1]Korea Advanced Institute of Science and Technology    [2]Samsung Electronics

2021.05.02

## Abstract

Anomaly detection in medical imaging is to distinguish the relevant biomarkers of diseases from those of normal tissues. Deep supervised learning methods have shown potentials in various detection tasks, but its performances would be limited in medical imaging fields where collecting annotated anomaly data is limited and labor-intensive. Therefore, unsupervised anomaly detection can be an effective tool for clinical practices, which uses only unlabeled normal images as training data. In this paper, we developed an unsupervised learning framework for pixel-wise anomaly detection in multi-contrast magnetic resonance imaging (MRI). The framework has two steps of feature generation and density estimation with Gaussian mixture model (GMM). A feature is derived through the learning of contrast-to-contrast translation that effectively captures the normal tissue characteristics in multi-contrast MRI. The feature is collaboratively used with another feature that is the low-dimensional representation of multi-contrast images. In density estimation using GMM, a simple but efficient way is introduced to handle the singularity problem which interrupts the joint learning process. The proposed method outperforms previous anomaly detection approaches. Quantitative and qualitative analyses demonstrate the effectiveness of the proposed method in anomaly detection for multi-contrast MRI.

1. Introduction

Detection and segmentation of medical images are useful tools for finding biomarkers that distinguish normal tissue characteristics from abnormalities. Recently, deep learning approaches have been developed for several detection tasks and their data-driven markers have provided convincing results [1,2]. Supervised learning has shown outstanding performances in specified disease detection for medical images [3]. However, the supervised learning requires a large amount of annotated images.

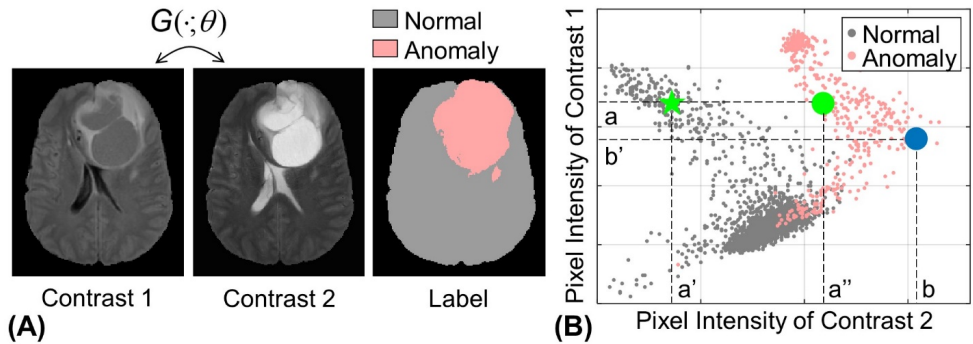

Figure 1. (A) Two MR images with different contrasts and a label image of anomaly region. (B) Intensity distribution of the two images. The green and blue points in (B) depicts the two cases of how anomaly can be detected in contrast-to-contrast translation.

Unsupervised anomaly detection is gaining interest in medical imaging fields [4]. Instead of using annotated anomaly images as references, the unsupervised approaches estimate the distributions of easily accessible normal images and distinguish exceptional data from large normal datasets [5]. The methods could be grouped into two representative categories: reconstruction-based [6-8] and clustering-based methods [9]. The reconstruction-based methods usually employ a deep autoencoder, assuming that the deep autoencoder trained by only normal images cannot properly reconstruct anomaly images. Meanwhile, the clustering-based methods use neural networks to extract latent features from high dimensional images, and the clustering algorithms such as One-Class Support Vector Machine (OC-SVM) [10] classify the features. These methods have provided promising results for the images of magnetic resonance imaging (MRI) and optical coherence tomography (OCT). However, the reconstruction-based methods have the difficulty of choosing the degree of dimension reduction [6]. The clustering-based methods have the separated pipelines of feature estimation and clustering. The approaches could easily lead to suboptimal performance because the first step of feature estimation is performed without information from the subsequent clustering models [9].

A set of multi-contrast images are usually obtained for accurate diagnosis in MRI. Figure 1 shows examples of the different contrast MR images and their anomalous pixel intensity values. The pixel intensities of anomalies can be distinguished from normal intensities in the two-dimensional contrast space, while the normal and abnormal intensities are significantly overlapped in each dimension (Fig. 1(B)). Actually, many clinical protocols for diagnosis consist of multi-contrast images. The aspect is well reflected in several public datasets, such as TCIA (The Cancer Imaging Archive) [11], ISLES (Ischemic Stroke Lesion Segmentation) [12] and BraTS (Brain Tumor Segmentation) [13]. Several anomaly detection approaches utilizing the inherent properties of multi-contrast images have been developed for detecting epilepsy or multiple sclerosis lesions [6,14-17].

In this work, a new unsupervised learning framework is proposed for pixel-wise anomaly detection in multi-contrast MR images, which is called the ADM network. The ADM network can effectively estimate the density of normal data in an unsupervised manner and achieve significant improvement in anomaly detection thanks to several contributions. First, two relevant features are extracted from multi-contrast MRI. One is derived through learning of contrast-to-contrast translation between multi-contrast MR images, and the other is the low-dimensional representation of multi-contrast images. Experimental results demonstrate that the contrast translation is an effective way to capture the accurate distribution of normal data, which is distinct from that of anomaly data. Second, the two different types of features are collaboratively used in a following density estimation model for Gaussian

Mixture Model (GMM). Also, the two steps of generating features and estimating density are jointly optimized so that the generated features become suitable for the anomaly detection tasks. Third, the proposed density estimation model effectively prevents the singularity problem that frequently occurs in estimating GMM parameters and critically interrupts the joint learning processes. The proposed method was compared with previous anomaly detection algorithms, in terms of how accurately anomaly pixels can be detected in multi-contrast MR images. Qualitative and quantitative analyses demonstrate the superiority and the feasibility of the proposed method in pixel-wise anomaly detection.

*1.1. Related Works*

A typical approach for unsupervised anomaly detection is to use reconstruction errors, assuming that anomalies cannot be effectively reconstructed by models representing normal data distributions. Thus, the degree of errors is the criterion for determining whether the data is normal or not. Most of the methods assume that there are great losses in the dimension reduction of anomalies when the process is optimized only for normal data. In this respect, several works have employed principal component analysis (PCA) which is a typical unsupervised dimension reduction algorithm [18,19]. Recently, a deep autoencoder has shown remarkable performance [20-22]. The deep autoencoder compresses input data into low dimensional latent variables in the encoder part. In training, the reduced latent variables represent the common features of training datasets, from which a decoder accurately reconstructs the original input data. So, the deep autoencoder trained only with normal datasets cannot properly reconstruct anomalies [6-8].

However, it is difficult to determine the degree of dimension reduction, which is an important hyper-parameter in determining reconstruction errors. The dimensions of latent variables should be properly selected to include the common features of normal datasets while excluding the features of anomalies. However, it is usually difficult to estimate the intrinsic dimensionality for complex data [23]. If a reconstruction model compresses high dimensional input data into few variables (for example, an eigenvector in PCA or a latent value in an autoencoder), this model poorly reconstructs both normal and anomaly data, so that it cannot be appropriate to detect anomalies. On the other hand, if we choose a too high dimension, even anomaly data can be reconstructed well.

Another common method for anomaly detection is to use a clustering algorithm to estimate the probability distribution of normal data. Conventional clustering and classification algorithms including kernel density estimation, support vector machine and GMM have been used in anomaly detection [24-27]. However, it is known that these algorithms are easy to fail especially in high-dimensional datasets because of the curse of dimensionality [26]. Therefore, deep neural networks have been adopted for dimension reduction as effective latent representations, and the reduced latent variables were used as the inputs of anomaly detection models [9,14,28-30]. These deep networks have shown promising results, but the performances of these methods were limited by the fact that the latent features were not suitable for following clustering models.

To overcome the issues, a research group proposed a clustering objective function for a neural network [31], which was based on support vector data description [32]. The latent variables induced by the objective function were directly used for anomaly detection without clustering models. In another way, some groups have developed joint learning of dimension reduction and clustering models [33-35]. In the joint learning frameworks, neural networks estimated latent representations suitable for clustering models (K-means or GMM). Especially, the joint learning framework with GMM showed outstanding performance in an anomaly detection field [36]. However, estimating the parameters of GMM in the joint learning model may introduce the singularity problem. The singularity problem is that a variance of the Gaussian model becomes zero in a specific dimension so that its covariance matrix becomes singular (or non-invertible). The singular matrix critically interrupts the learning processes of estimating GMM parameters [36,37].

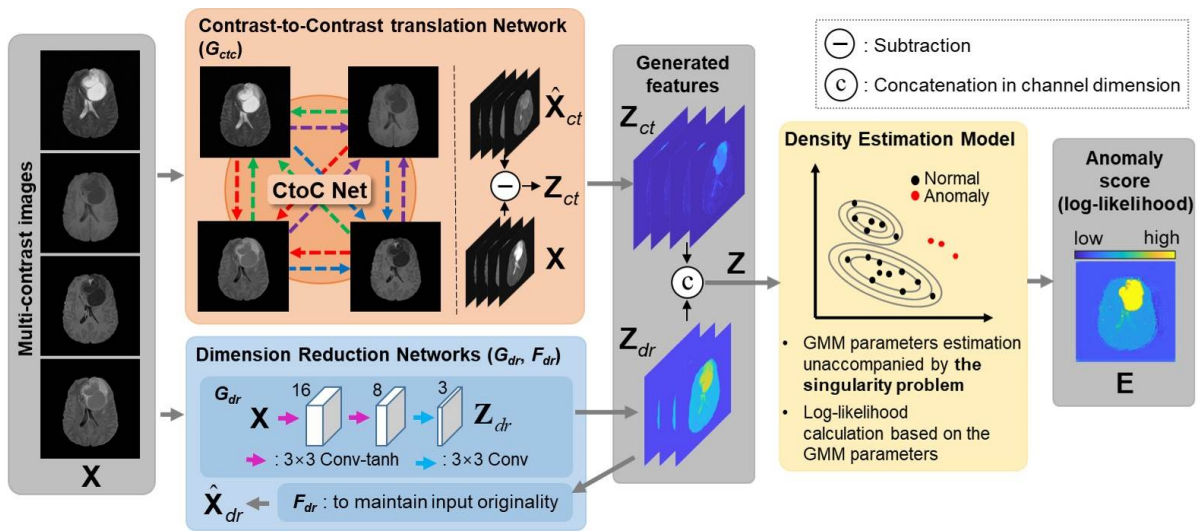

Figure 2. Overall diagram of the ADM network in the test phase. Two different types of features are extracted from multi-contrast MR images, which are reconstruction-based (the orange box) and clustering-based (the blue box), respectively. The dotted arrows in the CtoC network denote contrast translation between multi-contrast images. In a density estimation model (the yellow box), normative distributions estimated in training phases are used to detect anomalies. The region considered anomalies has high intensity in **E**.

Recently, powerful deep neural networks have been developed for the general representation learning [38,39]. By using the neural networks, several anomaly detection techniques have shown promising results. First, a variational autoencoder (VAE) was one of popular neural networks adopted for unsupervised anomaly detection [6,16,40], which used probabilistic latent spaces to describe normative distribution effectively [41]. Second, a generative adversarial network (GAN) [39] have been recently used for anomaly detection techniques [42-44]. Some of these networks combined an autoencoder with GAN for relevant latent representations [7]. In the deep framework, the latent space of normal images was directly guided by GAN [45]. The VAE-based and GAN-based methods improved anomaly detection performance. However, since their architectures are based on autoencoders, the difficulty of choosing the reduction degree still remains.

## 2. Materials and Methods

A new anomaly detection algorithm is proposed for multi-contrast MR images. The proposed framework, which is called the ADM network, has two steps of feature generation and density estimation of the given features (Fig. 2). Two different types of features are collaboratively used for representing normal data density and detecting anomalies. The motivation of each module is explained in detail as follows.

## 2.1. Motivations of the Proposed Method

A feature for the proposed anomaly detection ($Z_{ct}$) is derived from the contrast relationship between multi-contrast images. Learning of contrast-to-contrast translation is a way of generating a specific contrast image from the other contrast images. A network trained with only normal data can effectively capture the relationship of multi-contrast normal images. In the test phase, a specific contrast image is generated from the other contrast images by the trained network. Then, the reconstruction error between the generated image and the true image is the basis for detecting anomalies.

Figure 1(B) shows an example of how the contrast translation works for anomaly detection. The green points of the star and circle in Fig. 1(B) depict the case when normal and anomaly tissues have similar intensities ($a$) in contrast 1 but significantly different ($a'$ and $a''$, respectively) in contrast 2. A neural network trained with only normal data for contrast translation from 1 to 2 poorly reconstructs the intensity of contrast 2 of the green circle, and its reconstruction error would be the difference between the two green points of the star and circle in the dimension of contrast 2 (i.e. $|a' - a''|$). The blue point in Fig. 1(B) describes another case where anomaly data has a completely different intensity ($b$) from normal data distribution in contrast 2. A neural network for contrast translation from 2 to 1 would have significant reconstruction errors on the unseen input data of blue circle. The reconstruction errors are derived using the relationship between multi-contrast images. That is, the proposed method is not based on the data compressibility of anomaly, unlike previous reconstruction-based methods using autoencoders. The effectiveness of the proposed reconstruction-based method is validated in our experiments.

Another type of feature ($Z_{dr}$) is the low-dimensional representation of multi-contrast images. Normal intensities are distributed differently from anomaly intensities in multi-contrast images (Fig. 1(B)). To extract relevant features from the multi-contrast images, a neural network is adopted for dimension reduction. The trained network with only normal multi-contrast images would represent a low-dimensional latent space for normal data so that anomalies could have different distributions in the latent space. Note that the joint learning scheme is used to overcome the issue that the estimated features would be less effective for clustering [28,30]. That is, the neural network for dimension reduction and the following model for clustering are jointly optimized in the proposed method. The effectiveness of the joint learning is demonstrated through experiments.

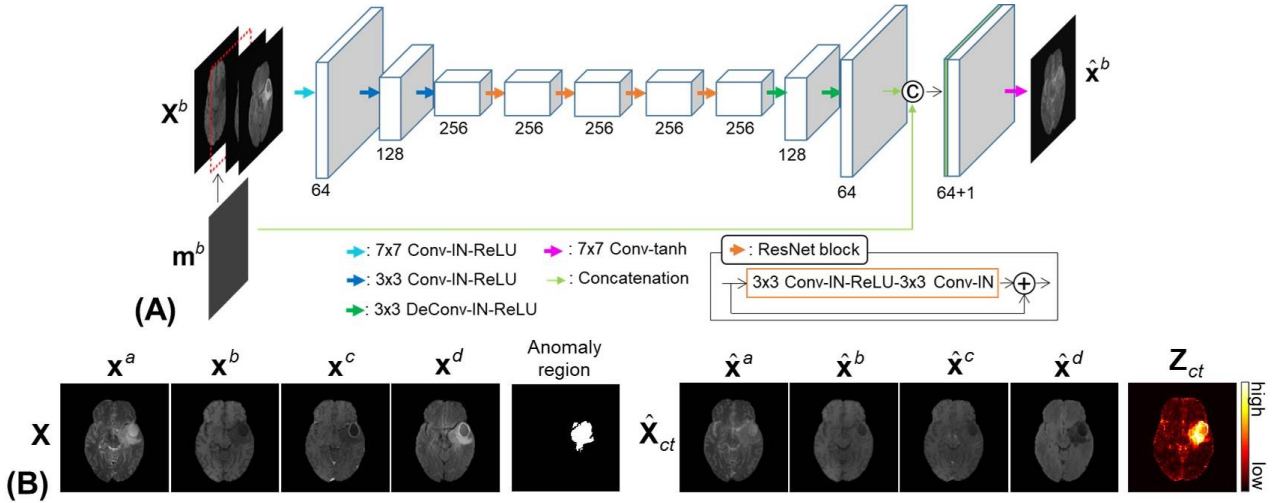

Figure 3. (A) The architecture of the CtoC network. The $\mathbf{m}^b$ is fed into the first and the last layers in the CtoC network to provide an intensity bias in synthesizing a target contrast image. (B) Examples of the input $\mathbf{X}$ and output $\hat{\mathbf{X}}_{ct}$ of the CtoC network and their anomaly region. For visualization, the reconstruction errors $\mathbf{Z}_{ct}$ is averaged along the contrast dimension.

A density estimation model is adopted to properly describe normal feature distribution as GMM parameters. A log-likelihood function with the estimated GMM parameters is calculated for scoring the anomaly degree. However, estimating GMM parameters is vulnerable to the singularity problem when a covariance matrix becomes singular, and learning processes are interrupted due to the singularity problem. Even though several approaches have been developed to prevent this singularity problem [36,37], these approaches would be effective only if the feature distributions do not change during density estimation. Therefore, we propose a density estimation model to prevent the singularity problem in the joint learning, by ensuring a positive definite covariance matrix.

## 2.2. Architectures of ADM Network

**Contrast-to-Contrast Translation (CtoC) network**. The CtoC network ($G_{ct}$) generates a specific contrast image from the collaborative knowledge of other contrast images (Fig. 3). Based on the architecture showing impressive results in style transfer [46], the CtoC network is designed to perform contrast translation with the prior information of a target contrast. First, we assume that there are a set ($\mathbf{X}$) of four contrast input images ($\mathbf{x}^a, \mathbf{x}^b, \mathbf{x}^c$ and $\mathbf{x}^d$) as follows.

$$\mathbf{X} = \{\mathbf{x}^a, \mathbf{x}^b, \mathbf{x}^c, \mathbf{x}^d\}$$

Assuming that an image of $b$ contrast is generated by $G_{ct}$, it can be described as follows.

$$\hat{\mathbf{x}}^b = G_{ct}(\mathbf{X}^b, \theta_{ct})$$

$$\text{where } \mathbf{X}^b = \{\mathbf{x}^a, \mathbf{m}^b, \mathbf{x}^c, \mathbf{x}^d\},$$

$\theta_{ct}$ is the parameters of the network $G_{ct}$, and $\hat{\mathbf{x}}^b$ and $\mathbf{X}^b$ are the output and the input of the CtoC network, respectively. Here, $\mathbf{m}^b$ is an image whose whole pixels are filled with the median value of $\mathbf{x}^b$. The median value as prior information serves to consider the intensity diversity of open datasets. A deep neural network is trained for all cases of contrast translation. The CtoC network uses the channel location of m in input data to identify a translation case among all cases. That is, a generated image has the contrast corresponding to the channel location of m. For example, when m is the image of a channel $b$, the network generates a $b$ contrast image. Reconstruction errors between the generated image and the corresponding true image are used as features ($\mathbf{Z}_{ct}$) for anomaly detection in a test phase. Examples of the generated images, $\hat{\mathbf{X}}$, and their $\mathbf{Z}_{ct}$ are shown in Fig. 3(B) and their equations can be written as follows.

$$\begin{aligned}\hat{\mathbf{X}}_{ct} &= \{\hat{\mathbf{x}}^a, \hat{\mathbf{x}}^b, \hat{\mathbf{x}}^c, \hat{\mathbf{x}}^d\} \\ &= \{G_{ct}(\mathbf{X}^a; \theta_{ct}), \ldots, G_{ct}(\mathbf{X}^d; \theta_{ct})\}\end{aligned}$$

$$\mathbf{Z}_{ct} = abs(\mathbf{X} - \hat{\mathbf{X}}_{ct})$$

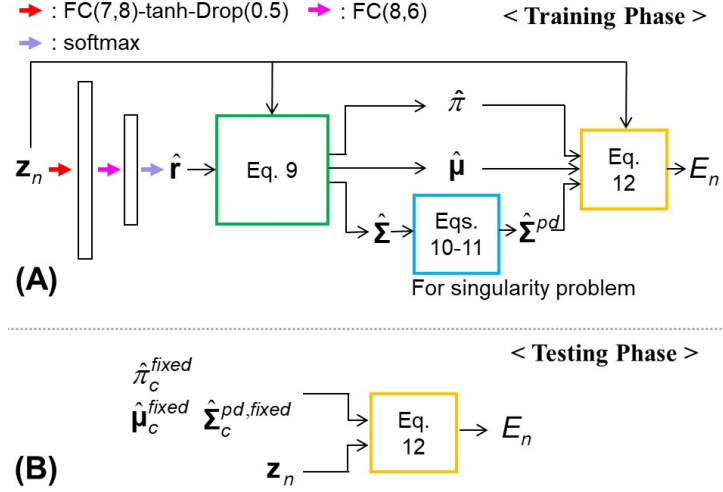

Figure 4. The architecture of the density estimation (DE) model. In the training phase (A), the DE model is trained to estimate the GMM parameters of given features $\mathbf{z}_n$ by reducing the log-likelihood $E_n$. In testing phase (B), the DE calculates the $E_n$ of $\mathbf{z}_n$ with $\hat{\pi}_c^{fixed}$, $\hat{\boldsymbol{\mu}}_c^{fixed}$ and $\hat{\boldsymbol{\Sigma}}_c^{pd,fixed}$, which are the GMM parameters estimated from the training data at the end of the training phase.

where abs(**X**) is the absolute value of the elements of **X**.

**Dimension Reduction (DR) Network**. The DR network ($G_{dr}$) generates features, $\mathbf{Z}_{dr}$, that correspond to the low dimensional representations of input images. Dimension reduction is performed along the channel dimension, whereas the spatial dimension of $\mathbf{Z}_{dr}$ is the same as that of input images. That is, each pixel intensity in $\mathbf{Z}_{dr}$ is a feature used in determining whether the pixel is anomaly or not. A reconstruction network ($F_{dr}$) is adopted to prevent $\mathbf{Z}_{dr}$ from losing key information of input X. The two networks are described in Fig. 2 and formulated as follows.

$$\mathbf{Z}_{dr} = G_{dr}(\mathbf{X}; \theta_{dr,1})$$

$$\hat{\mathbf{X}}_{dr} = F_{dr}(\mathbf{Z}_{dr}; \theta_{dr,2})$$

where $\theta_{dr,1}$ and $\theta_{dr,2}$ are the network parameters of $G_{dr}$ and $F_{dr}$, respectively, and $\hat{\mathbf{X}}_{dr}$ is the reconstructed images from $\mathbf{Z}_{dr}$.

The convolutional layers in $G_{dr}$ estimate the pixel-level features $\mathbf{Z}_{dr}$ by considering spatial information in a receptive field. In $F_{dr}$, 1x1 convolutional layers are used to offer a strong constraint that $\mathbf{Z}_{dr}$ maintains the spatial specificity of original input **X**. The four multi-contrast images are converted to three feature images. In training, $G_{dr}$ and $F_{dr}$ are jointly optimized with a following density estimation model.

**Density Estimation (DE) Model**. The DE model estimates GMM parameters well representing the normal features of training data and then calculates anomaly scores for test features by using the estimated GMM parameters (Fig. 4). The goodness of fit of the estimated GMM parameters is measured by a log-likelihood function. Log-likelihood is a loss function for training and also an anomaly score function for testing. Detailed configurations are as follows.

A vector along the channel dimension of the concatenation of $\mathbf{Z}_{ct}$ and $\mathbf{Z}_{dr}$ is a feature for pixel-wise anomaly detection, which is $\mathbf{z}_n$ ($1 \leq n \leq N$). Note that N is the total number of $\mathbf{z}_n$ in spatial and batch dimensions. The proposed DE model is based on previous works [36,47], where a neural network estimates the probability $\hat{r}_n$ that the input feature $\mathbf{z}_n$ belongs to each GMM. The GMM parameters of the fraction ($\hat{\pi}_c$), mean ($\hat{\boldsymbol{\mu}}_c$) and covariance ($\hat{\boldsymbol{\Sigma}}_c$) are calculated as follows.

$$\hat{r}_n = G_{de}(\mathbf{z}_n; \theta_{de})$$

$$\hat{\pi}_c = \sum_n \frac{\hat{r}_{n,c}}{N}, \hat{\boldsymbol{\mu}}_c = \frac{\sum_n \hat{r}_{n,c} \cdot \mathbf{z}_n}{\sum_n \hat{r}_{n,c}}$$

$$\hat{\boldsymbol{\Sigma}}_c = \frac{\sum_n \hat{r}_{n,c} (\mathbf{z}_n - \hat{\boldsymbol{\mu}}_c)(\mathbf{z}_n - \hat{\boldsymbol{\mu}}_c)^T}{\sum_n \hat{r}_{n,c}}$$

where $G_{de}$ and $\theta_{de}$ are the neural network and its parameters, respectively, and the subscript of c is an index for Gaussian models. The number of Gaussian models is a hyper-parameter $C$ and $\hat{r}_n$ has $C$ elements of $\hat{r}_{n,c}$ ($1 \leq c \leq C$). Six Gaussian models are assumed in our experiments.

The proposed DE model effectively prevents the singularity problem by introducing a lower bound ($\varepsilon$) for eigenvalues of a covariance matrix. The procedure ensures a covariance matrix to be positive definite as follows.

$$\hat{\boldsymbol{\Lambda}}_c^p = ReLU(\hat{\boldsymbol{\Lambda}}_c - \varepsilon) + \varepsilon$$

$$\hat{\boldsymbol{\Sigma}}_c^{pd} = \hat{Q}_c \cdot \hat{\boldsymbol{\Lambda}}_c^p \cdot \hat{Q}_c^T$$

where $\hat{Q}_c$ and $\hat{\boldsymbol{\Lambda}}_c$ are the eigenvectors and the eigenvalues of $\hat{\boldsymbol{\Sigma}}_c$, respectively, the superscript $T$ is the matrix transpose, and $\hat{\boldsymbol{\Sigma}}_c^{pd}$ is a final covariance matrix. The value of $\varepsilon$ was $10^{-6}$ in our experiments. Finally, the log-likelihood of $\mathbf{z}_n$ is calculated as follows.

$$E_n = -\log\left(\sum_c \hat{\pi}_c \frac{\exp(-0.5 \cdot (\mathbf{z}_n - \hat{\boldsymbol{\mu}}_c)^T (\hat{\boldsymbol{\Sigma}}_c^{pd})^{-1}(\mathbf{z}_n - \hat{\boldsymbol{\mu}}_c))}{\sqrt{\det(2\pi\hat{\boldsymbol{\Sigma}}_c^{pd})}}\right)$$

where $det(\cdot)$ denotes the determinant of a matrix. In the training phase, the DE model estimates GMM parameters for each batch data and minimizes their log-likelihood. That is, the DE model is optimized to be a density estimator. At the end of the training, final GMM parameters for normative distributions are estimated from the large amount of normal data. In the test phase, the estimated GMM parameters are used in calculating log-likelihoods for test data, which are anomaly scores.

Details of the parameters of the ADM network are explained in Section I of the Supplementary Materials.

### 2.3. Objective Function and Joint Learning Strategy

The proposed framework is optimized in two steps by using the two objective functions as follows.

$$\min_{\theta_{ct}} \|\mathbf{X} - \hat{\mathbf{X}}_{ct}\|_2^2$$

$$\min_{\theta_{dr,1},\theta_{dr,2},\theta_{de}} \|\mathbf{X} - \hat{\mathbf{X}}_{dr}\|_2^2 + \frac{\lambda}{N}\sum_n E_n$$

where $\|\cdot\|_2$ is the L2 norm of a vector. First, the $G_{ct}$ is optimized to minimize the objective function in Eq. 13, then the feature $\mathbf{Z}_{ct}$ is generated from $G_{ct}$. With the estimated $\mathbf{Z}_{ct}$, the $G_{dr}$ and the $G_{de}$ are jointly optimized to minimize the objective function in Eq. 14. The density estimation network finds the parameters of Gaussian mixtures, properly presenting the feature distributions by reducing $E_n$. As $E_n$ decreases, the features of $\mathbf{Z}_{dr}$ are densely distributed and have cluster-friendly properties. In the test phase, the features of anomalies are effectively discriminated by their $E_n$ values. The number of training epochs were 50 for BraTS dataset and 200 for ISLES dataset, respectively. The number of epochs was empirically determined based on the saturation of the loss values of Eqs. 13 and 14 (Fig. S1 in the Supplementary Materials). The $\lambda$ is a hyper-parameter for the ADM network. In practice, $\lambda$ was set as 0.0005 and 0.005 for BraTS and ISLES datasets, respectively, with which desirable results of the area under curve (AUC) were obtained using the validation data.

*2.4. Datasets*

Two open datasets of brain tumor segmentation (BraTS) 2019 [13] and ischemic stroke lesion segmentation (ISLES) 2015 [12] were used in the experiments. Both datasets are three-dimensional whole brain MR images with pixel-wise annotations of anomalies. Normal images and anomaly images were divided by using the provided annotation to perform anomaly detection experiments.

BraTS is a dataset for glioblastoma and lower grade glioma, and has the four contrast images of T2 FLAIR, T2-weighted, T1-weighted and T1-weighted with contrast agent. First, 241 patients in BraTS were randomly divided into two parts of 221 patients and 20 patients. From the 221 patients, 7790 normal image sets and 14185 anomaly image sets were used as training and validation data, respectively. The whole image sets from the 20 patients (2100 sets) were used as test data.

ISLES is a dataset for ischemic stroke lesions and has four contrast images such as diffusion weighted, T1 weighted, T2-FLAIR, and T2 weighted images. From 18 patients, 828 normal image sets and 900 anomaly image sets were selected and used as training and validation data, respectively. In additions, 875 image sets from other 6 patients were used as test data. Anomaly detection in ISLES was more challenging due to its small amount of data. In additional experiments, the training data were augmented with the images of the three common MR contrasts in BraTS and ISLES, which were Flair, T1 weighted and T2 weighted images.

The same training, validation and test datasets were used for the proposed and baseline models. Evaluating with the validation data can provide the results regarding not only abnormal tissues but also normal tissues, since anomaly images in validation data usually contain both normal and abnormal tissues.

Images from the two datasets went through the three preprocessing steps of skull stripping, co-registration between multi-contrast images and interpolation with $1mm^3$ resolution. The intensity distribution of whole brain images was normalized to be a normal distribution with zero mean and unit variance. This normalization was performed for each contrast image. All images were resized to $128 \times 128$ by using bilinear interpolation. Data augmentation with random scaling of the image intensity was performed.

*2.5. Quantitative Analyses*

Several metrics were adopted to quantify the performance of anomaly detection. The area under the curve (AUC) of the receiver operating characteristic (ROC) curve was measured to compute the general performance of anomaly detection. The best $F_1$ score for each approach was reported. A threshold value for the $F_1$ score was selected using the validation data. By using the threshold for the best $F_1$ score, precision and recall were also calculated to evaluate how accurately true positives were detected. Note that anomaly class was assumed as positive. Black background regions in brain images were not considered as normal class and excluded in calculating the metrics. The pixel-level annotations provided by each dataset were used as the ground truth.

*2.6. Baseline Models*

The proposed method was compared with several anomaly detection algorithms, which have been developed for general applications as well as medical images. For the three groups described in the Related Works, the representative works of the competing models were selected. In particular, an anomaly detection algorithm with GMM was selected and compared with the proposed method in terms of the singularity problem of GMM. The selected works are as follows. 1) One class support vector machine ($OCSVM$) [10] is a popular kernel-based technique for anomaly detection. 2) Gaussian mixture variational auto-encoder ($GMVAE$) is a reconstruction-based technique to detect anomalies in MRI, which is based on a variational autoencoder and a maximum-a-posterior (MAP) reconstruction [41]. 3) Deep support vector data description ($DSVDD$) [31] is a clustering-based anomaly detection algorithm introducing a clustering objective function. 4) $fanoGAN$ [45] used a new neural network with GAN for anomaly detection. Adversarial learning with GAN enables to secure the relevant latent representations of normal data. 5) Deep autoencoding GMM ($DAG$) [36] is an autoencoder adopting GMM for anomaly detection. 6) A modified DAG including convolutional layers is more appropriate for pixel-level anomaly detection. The modified method is called $DAG_c$. Similar to the $ADM$ network, hyper-parameter adjustment was performed using AUC values for the validation data, in order to obtain better performance for each method. The configuration of each method is described in Section III of the supplementary materials.

In addition, variants of the proposed framework were employed for the ablation study to demonstrate the importance of each component as follows. 7) $ADM_{ct}$ and 8) $ADM_{dr}$ are the proposed frameworks without the DR networks and the CtoC network, respectively. 9) $ADM_{woj}$ is the proposed framework without the joint optimization of feature generation and density estimation. That is, the two loss terms in Eq. 14 were used separately to optimize the corresponding network.

*2.7. Lower Bound for the Singularity Problem*

Density estimation using GMM has been used in various clustering tasks including anomaly detection [34,37,47]. The singularity problem is a known issue with GMM and happens more easily in the joint learning of feature generation and density estimation. We performed an experiment to investigate how the singularity problem occurred during the joint learning and to demonstrate the effectivness of the proposed lower bound method.

In this study, three types of the density estimation models were compared, which were the baseline model without any constraints, the proposed model using the lower bound for eigenvalues and the comparison model with the loss penalizing small diagonal entries in a covariance matrix [36]. For the comparison model, the term of $L_p = \sum_i^d \frac{1}{\hat{\Sigma}_{c,ii}}$ is added to the loss of Eq. 14 with $\lambda$ of $10^{-5}$, where $d$ is the number of diagonal entries. This method was the generalized version of adding a small value to the diagonal entries. The joint learning model of $ADM_{dr}$ was employed to visualize its three dimensional feature space.

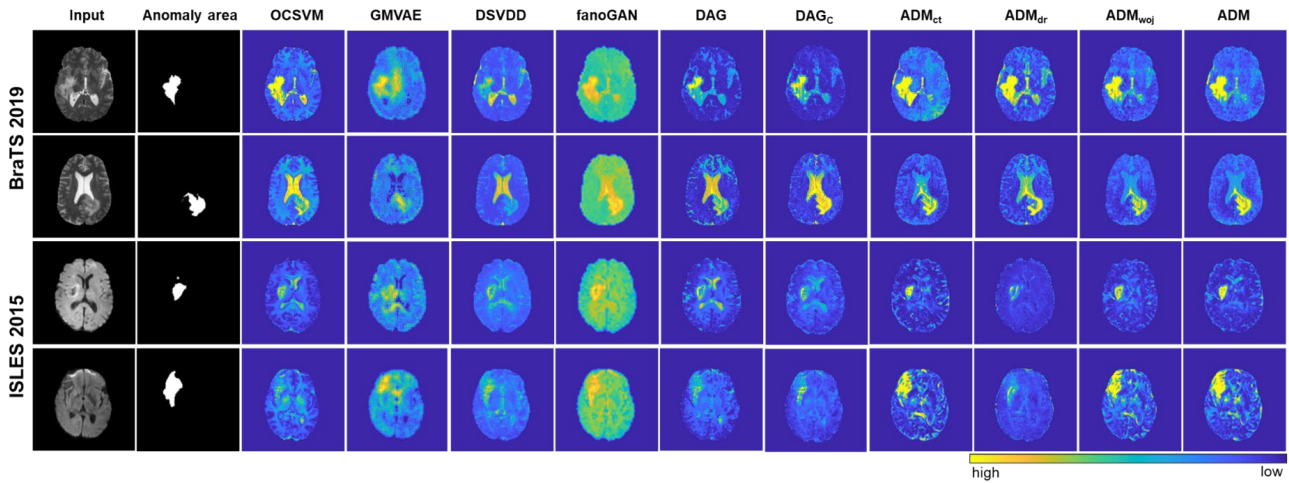

Figure 5. Resultant anomaly score maps ($E$) for test and validation data in BraTS 2019 and ISLES 2015. One of multi-contrast images is shown as an input image. High intensity values mean the high probability of anomaly.

## 3. Results

### 3.1. Quantitative Analyses

Quantitative analysis results are shown with AUC, Precision, Recall and F1 score for BraTS and ISLES (Table I,II,III and IV). Each method was trained three times and then the mean and the standard deviation of each metric were measured. Examples of anomaly score maps for each method are depicted as qualitative results (Fig. 5). The results for validation data usually show lower performance than those for test data, since validation data consisting of only anomaly images have a higher anomaly ratio than test data.

Anomaly detection in ISLES is more challenging than in BraTS, because of the small amount of training data in ISLES. The quantitative values in ISLES are usually lower than those in BraTS. Lower precision values are observed in the test data, compared to those in the validation data, since insufficient normal data cannot represent the general properties of normal tissues. Nevertheless, the proposed ADM method still shows higher performance than the competing models.

***Comparison Study.*** Table I and II show the quantitative results for comparison study with the mean and the standard deviation of each metric. The $F_1$ score of the *ADM* is higher than those of the competing models (BraTS validation/test = 0.570/0.683, ISLES validation/test = 0.413/0.278). In addition, the *ADM* also achieves the highest AUC value for BraTS validation/test data (0.880/0.938) and ISLES test data (0.871), while *DSVDD* provides a higher AUC value for ISLES validation data and the *ADM* is ranked 2nd. For the precision values, the proposed method is usually ranked high, except for in the ISLES validation data, as follows: BraTS validation/test

= 1st/2nd and ISLES validation/test = 5th/1st. For recall values, the ranks of the proposed method are as follows: BraTS validation/test = 4th/2nd, ISLES validation/test = 2nd/1st.

The AUC values show general performance in anomaly detection. For the AUC values of test data, a paired t-test analysis was performed between ADM and competing methods. In the statistical analyses, $p$ values were below 0.05 for most cases, except $DAG_c$ for BraTS ($p$ = 0.08). Consequently, the ADM method usually achieved higher performance in anomaly detection. These quantitative results are well reflected in anomaly score maps. The proposed method shows low anomaly scores in most normal regions (Fig. 5) and it leads to low false positives and high performance. In contrast, some competing models provide high anomaly scores for normal tissues, especially around cerebrospinal fluid (CSF).

Table I. Quantitative results in comparison studies for BraTS datasets

|  |  | OCSVM | GMVAE | DSVDD | fanoGAN | DAG | DAGc | ADM |
|---|---|---|---|---|---|---|---|---|
| Valid. | AUC | 0.725 ±0.000 | 0.759 ±0.020 | 0.826 ±0.006 | 0.764 ±0.000 | 0.836 ±0.001 | 0.843 ±0.027 | **0.883 ±0.002** |
|  | Precision | 0.401 ±0.000 | 0.255 ±0.051 | 0.425 ±0.023 | 0.426 ±0.014 | 0.430 ±0.029 | 0.449 ±0.062 | **0.578 ±0.042** |
|  | Recall | 0.404 ±0.000 | 0.519 ±0.101 | 0.597 ±0.036 | 0.392 ±0.011 | 0.587 ±0.049 | **0.618 ±0.040** | 0.574 ±0.021 |
|  | $F_1$ | 0.402 ±0.000 | 0.334 ±0.029 | 0.496 ±0.003 | 0.408 ±0.000 | 0.495 ±0.002 | 0.519 ±0.056 | **0.575 ±0.010** |
| Test | AUC | 0.818 ±0.000 | 0.790 ±0.036 | 0.866 ±0.004 | 0.832 ±0.000 | 0.900 ±0.001 | 0.909 ±0.015 | **0.939 ±0.001** |
|  | Precision | 0.588 ±0.000 | 0.335 ±0.023 | **0.727 ±0.057** | 0.638 ±0.011 | 0.581 ±0.006 | 0.602 ±0.094 | 0.711 ±0.039 |
|  | Recall | 0.506 ±0.000 | 0.410 ±0.084 | 0.574 ±0.030 | 0.432 ±0.005 | 0.653 ±0.019 | **0.691 ±0.007** | 0.675 ±0.024 |
|  | $F_1$ | 0.544 ±0.000 | 0.366 ±0.037 | 0.640 ±0.004 | 0.515 ±0.000 | 0.615 ±0.005 | 0.641 ±0.050 | **0.692 ±0.006** |

Table II. Quantitative results in comparison studies for ISLES datasets

|  |  | OCSVM | GMVAE | DSVDD | fanoGAN | DAG | DAGc | ADM |
|---|---|---|---|---|---|---|---|---|
| Valid. | AUC | 0.686 ±0.000 | 0.716 ±0.010 | **0.846 ±0.004** | 0.730 ±0.000 | 0.731 ±0.003 | 0.736 ±0.017 | 0.826 ±0.003 |
|  | Precision | 0.455 ±0.000 | 0.256 ±0.045 | 0.344 ±0.006 | **0.569 ±0.009** | 0.541 ±0.010 | 0.437 ±0.087 | 0.435 ±0.023 |
|  | Recall | 0.215 ±0.000 | 0.227 ±0.008 | **0.520 ±0.015** | 0.225 ±0.002 | 0.241 ±0.005 | 0.276 ±0.016 | 0.385 ±0.020 |
|  | $F_1$ | 0.292 ±0.000 | 0.239 ±0.018 | **0.413 ±0.008** | 0.322 ±0.000 | 0.334 ±0.003 | 0.335 ±0.019 | **0.413 ±0.002** |
| Test | AUC | 0.593 ±0.000 | 0.637 ±0.019 | 0.609 ±0.008 | 0.665 ±0.000 | 0.642 ±0.005 | 0.656 ±0.018 | **0.871 ±0.004** |
|  | Precision | 0.057 ±0.000 | 0.057 ±0.005 | 0.049 ±0.002 | 0.167 ±0.006 | 0.081 ±0.003 | 0.075 ±0.029 | **0.205 ±0.007** |
|  | Recall | 0.393 ±0.000 | 0.337 ±0.012 | 0.391 ±0.034 | 0.226 ±0.005 | 0.280 ±0.031 | 0.366 ±0.134 | **0.429 ±0.007** |
|  | $F_1$ | 0.098 ±0.000 | 0.097 ±0.008 | 0.087 ±0.003 | 0.192 ±0.006 | 0.125 ±0.005 | 0.116 ±0.024 | **0.278 ±0.005** |

Table III. Quantitative results in ablation studies for BraTS

|  |  | $ADM_{ct}$ | $ADM_{dr}$ | $ADM_{woj}$ | ADM |
|---|---|---|---|---|---|
| Valid. | AUC | 0.861 | 0.865 | 0.878 | **0.883** |
|  | Precision | 0.527 | **0.623** | 0.576 | 0.578 |
|  | Recall | 0.522 | 0.550 | 0.569 | **0.574** |
|  | $F_1$ | 0.521 | **0.584** | 0.572 | 0.575 |
| Test | AUC | 0.922 | 0.916 | 0.934 | **0.939** |
|  | Precision | 0.678 | **0.771** | 0.711 | 0.711 |
|  | Recall | 0.580 | 0.632 | 0.637 | **0.675** |
|  | $F_1$ | 0.625 | **0.693** | 0.678 | 0.692 |

Table IV. Quantitative results in ablation studies for ISLES

|  |  | $ADM_{ct}$ | $ADM_{dr}$ | $ADM_{woj}$ | ADM |
|---|---|---|---|---|---|
| Valid. | AUC | 0.772 | 0.737 | 0.788 | **0.826** |
|  | Precision | 0.283 | **0.610** | 0.331 | 0.435 |
|  | Recall | 0.351 | 0.257 | 0.355 | **0.385** |
|  | $F_1$ | 0.313 | 0.362 | 0.342 | **0.413** |
| Test | AUC | 0.835 | 0.674 | 0.847 | **0.871** |
|  | Precision | 0.128 | 0.100 | 0.153 | **0.205** |
|  | Recall | **0.524** | 0.308 | 0.475 | 0.429 |
|  | $F_1$ | 0.205 | 0.148 | 0.232 | **0.278** |

The $OCSVM$ provides the lowest performance in the AUC and $F_1$. The $GMVAE$ has lower precision than the $ADM$ and it leads to low performance in AUC and $F_1$. The $DSVDD$ usually shows high recall and low precision. For ISLES validation data, the $DSVDD$ has higher AUC and recall values than the $ADM$. The $fanoGAN$ provides relatively high anomaly scores in normal regions, but sufficient differences between normal and anomaly regions can be observed in anomaly score maps (Fig. 5). The $DAG_c$ that is the convolutional variant of $DAG$ usually outperforms $DAG$ in most evaluations and is second only to the $ADM$ for BraTS.

*Ablation Study.* Table III and IV show the quantitative results for ablation study with the mean of each metric. For BraTS, the AUC values of $ADM_{ct}$ (validation/test = 0.861/0.922) are comparable to those of $ADM_{dr}$ (validation/test = 0.865/0.916). However, for the small dataset of ISLES, the AUC values of $ADM_{ct}$ (validation/test = 0.772/0.835) are higher than those of $ADM_{dr}$ (validation/test = 0.737/0.674). The $ADM_{woj}$ that is a combination of $ADM_{ct}$ and $ADM_{dr}$ outperforms each variant for AUC values. In addition, the joint learning in the proposed method provides further improvement, given that $ADM$ usually shows better performance in the quantitative results than $ADM_{woj}$. For the trained $ADM$ model, we reconstructed two anomaly score maps from the two types of

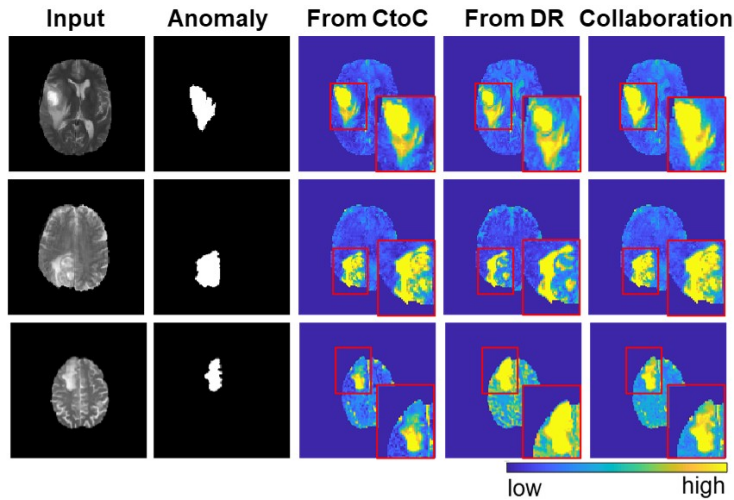

Figure 6. Anomaly score maps estimated from three types of features in ADM model.

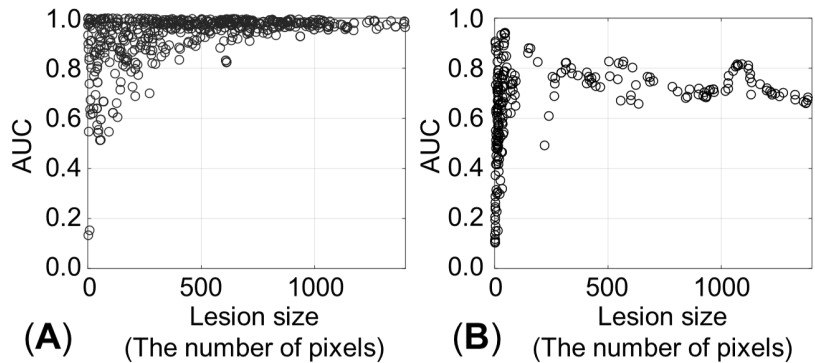

Figure 7. AUC values with respect to anomaly size for BraTS test data (A) and ISLES test data (B). An AUC value was measured for a slice. For visualization, 500 AUC values were randomly selected and depicted in the figures.

features which are induced from the CtoC and DR network, respectively, and compared with the anomaly score maps of the original *ADM* model (Fig. 6)

*Analyses with Anomaly Size.* We analyzed the performance of the *ADM* in terms of anomaly size. The anomaly size was measured as the number of annotated lesion pixels in an image, and the AUC values for various anomaly sizes are shown in Fig. 7.

### 3.2. Lower Bound for the Singularity Problem

Figure 9(A,B) describes the features **Z** of the $ADM_{dr}$ without any constraints (the baseline model), just before the singularity problem occurs, where the two figures describe the same feature space but have different view angles. The features were properly described by the Gaussian model but collapsed to certain dimension. That is, the features were placed on a two-dimensional plane (depicted as red color in Fig.9(B)) and the direction perpendicular to the red plane was the collapsed dimension. Consequently, the Eigen value of a covariance matrix along the dimension was close to zero (three Eigen values were $3.1 \times 10^{-7}$, 0.0021 and 0.0166) and the covariance

matrix became a singular matrix (non-invertible). The reason why the singularity problem occurs is that the dimension of feature space is higher than the intrinsic dimension of features. However, it is difficult to find the intrinsic dimension of latent features of inputs [23]. So, the singularity problem would be inevitable.

We compared the three DE models by visualizing the log-likelihood loss of Eq.14 (Fig. 8(c)). The ends of the curves of 'Baseline' and 'Ref. [36]' mean the interruption of training due to the singularity problem. The baseline DE model without any constraints was easily interrupted (the blue line). Even the DE model with the loss in 'Ref. [36]' also falls in interruptions (the orange line). Furthermore, since the loss directly affected the log-likelihood value, the log-likelihood loss was slowly minimized compared to that of 'Baseline'. In contrast, the proposed DE model effectively prevented the singularity problem while the intended minimization of log-likelihood was not disturbed (the yellow line).

## 4. Discussions

In this study, we developed an unsupervised algorithm for pixel-level anomaly detection (i.e., anomaly segmentation) in multi-contrast MRI, based on combination of feature generation and density estimation with GMM. Two relevant features were extracted from multi-contrast MR images and collaboratively used for unsupervised anomaly segmentation. The three components of the ADM network, which were CtoC, DR and DE models, were well trained to do their corresponding tasks correctly (Fig. S1 in the Supplementary Materials). To evaluate the performance of the proposed method, six previous anomaly detection algorithms were compared in quantitative and qualitative comparison study. The proposed method demonstrated high performance in anomaly segmentation for glioblastoma and ischemic stroke lesion, compared to the previous approaches (Tables I and II). The CtoC network generated relevant reconstruction-based features by using the contrast relationship between multi-contrast images and outperformed the autoencoder-based works. To validate the effectiveness of each proposed module, we performed an ablation study by removing each module from the proposed framework (three variants of the proposed method were used). The quantitative results in Tables III and IV demonstrate the improvement gained as each module is added. Furthermore, the proposed method using the lower bound for the eigenvalues of covariance matrices effectively prevented the singularity problem in joint learning with GMM.

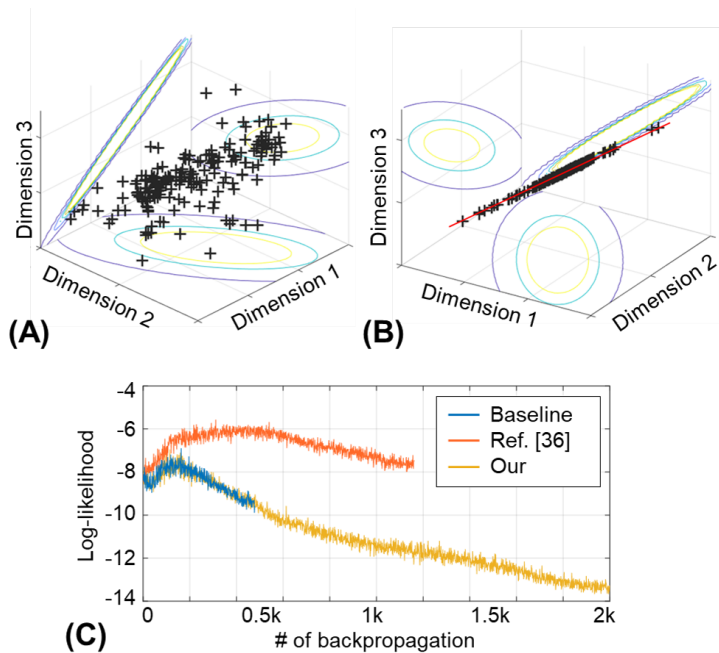

Figure 8. (A,B) Features and their latent spaces in the $ADM_{dr}$ are visualized with different view angles. (C) The evolution of log-likelihood loss for the three DE models.

The proposed ADM model adopted a contrast translation (CtoC) network to generate the reconstruction-based features. The effectiveness of the CtoC network was shown in the experimental results in Fig. 5 and Table I and II. Especially for ISLES, the competing models had a difficulty in generalizing normal tissue characteristics due to limited training data and showed low performance for test data (AUC < 0.8 and $F_1$ < 0.2 in Table II). In contrast, $ADM_{ct}$ showed higher performance for ISLES test, compared to competing models (AUC: 0.871 and $F_1$: 0.278) and led to the high performance of the ADM. Furthermore, the low-dimensional representation from the dimension reduction (DR) network could be another type of features for anomaly detection in multi-contrast MRI and could be collaboratively used with the features from the CtoC network. The ablation study demonstrated their complementary roles in quantitative ways. Also, the collaboration of the CtoC and the DR networks in the $ADM$ model are shown well in the anomaly score maps (Fig. 6).

Considering the high AUC and $F_1$ scores in the $ADM$ model, the proposed method generally showed superior performance to the competing methods for segmenting glioblastoma and ischemic stroke lesions. Low anomaly scores in most of normal regions and small false positives resulted in high precision of the proposed method, although some methods provided higher recall values. For $OCSVM$, the high dimensions of the image data limited its performance. For $GMVAE$, $fanoGAN$, $DAG$ and $DAG_c$ employing autoencoders for feature generation, the difficulty of finding the proper rate of dimension reduction would limit their performances. Although $DSVDD$ was comparable with the $ADM$ for ISLES validation, it had low performance for ISLES test. Test data would have new types of normal tissues, which are not included in train data. The $DSVDD$ seems to have a difficulty in generalizing normative distributions for ISLES data. That is why some normal regions, especially CSF, have higher energy than anomaly regions (Fig 5).

We analyzed the quantitative results of the ADM network according to the lesion size (Fig. 7). The AUC value usually decreases as the size of a lesion decreases. The ADM network shows stable performance for lesions of various size in BraTS. On the other hand, in ISELS dataset which has insufficient training data, the AUC values are relatively low especially for the small size of lesions. Furthermore, although the overall range of the lesion size of ISLES is similar to that of BraTS, ISLES has a lot of small size lesions. That would be a reason why anomaly detection in ISLES was more challenging.

The singularity problem is inevitable in joint learnging with GMM, as shown in Fig. 8(A,B). Reducing the likelihood of features in joint learning helps a neural network generate latent features to be suitable for Gaussian models. However, if the given dimension of the latent space is larger than the intrinsic feature dimension, the features collapse into the surplus latent dimension during training and the singularity problem occurs (Fig. 8(B)). It is difficult to find the intrinsic dimension of latent features of inputs [23]. Therefore, instead of challenging network design for an appropriate latent space, using a lower bound for the eigenvalue of covariance matrices fundamentally prevents the singularity problem. We demonstrated the efficacy of the proposed method in the comparison study (Fig. 8(C)). The method is expected to be applicable to various applications with GMM.

BraTS and ISLES were reconfigured to perform unsupervised anomaly detection experiments, which were originally intended for supervised learning. The proposed method demonstrated the possibility of detecting unspecified anomalies in the two datasets, although a large amount of normal MRI data were not fully utilized in training especially for ISLES. The straightforward experiments with augmented datasets for ISLES (section IV in the Supplementary Materials) demonstrated that increase of normal data could improve the performance of anomaly detection and verified the advantage of data accessibility to normal data. Nevertheless, using a lot of open and normal MRI datasets in [48] and [49] should be considered to further improve performance and be applicable to clinical practices.

The ADM network has shown reliable performance on open datasets of brain lesion, whose images were collected from various institutes. There have been problems with applicability to various clinical datasets, since the contrasts of MR images may vary according to clinical institutes even with the same MR protocols. To resolve the problems, we practically used two strategies. First, the CtoC network used the median value of a target contrast image (**m** in Eq. 3) as an input. Contrast translation by the CtoC network was not a deterministic mapping so that reconstruction errors can occur even with normal images. The median value provided an intensity bias in synthesizing the target contrast image. Second, data augmentation with random scaling of image intensity was also adopted to make the ADM network to be trained using datasets with the normative distributions of diverse contrasts. The data augmentation prevented the ADM network from overfitting to specific contrasts. As shown in Section V of the Supplementary Materials, further experiments were performed using the variants of the ADM network with or without both strategies. The AUC values for BraTS test were measured for evalutation (Table S3). The results show the effectiveness of the two strategies.

There are further works for the ADM network. First, there are some hyper-parameters that affect the performance of the ADM network (Section II of the Supplementary Materials). To obtain better performance, AUC values for validation data were used for adjusting the values of $\lambda$ (Eq. 14) of the ADM network. Therefore, the hyper-parameter selection for the unsupervised models was performed by using label information in the validation data. A new hyper-parameter selection method without label information should be developed for fully unsupervised learning and it would be a future work. Second, the usefulness of the ADM network was proven in its segmentation performance with BraTS and ISLES, where brain lesions were large and relatively evident. Evaluation to detect or segment small and subtle brain lesions, such as multiple sclerosis and vascular diseases, still remains.

## 5. Conclusions

We proposed an unsupervised anomaly detection algorithm for MRI, which extracted two relevant features from multi-contrast images and found anomaly regions by using a Gaussian mixture model. The proposed method has advantages in terms of pixel-wise detection performance and a learning strategy with GMM. The experimental results show that the proposed approach has higher performance than the state-of-the-art anomaly detection approaches. In particular, by forcing an estimated covariance matrix to be positive definite, the singularity problem can be prevented in GMM. The unsupervised approach has great potential in detecting various lesions in multi-contrast MRI.

# SUPPLEMENTARY MATERIALS

## I. CONFIGURATIONS OF THE ADM NETWORK

### A. Contrast-to-Contrast Translation (CtoC) Network

The instance normalization (IN) [1] and ReLU [2] were used. The stride in the 3×3 Conv and DeConv layers was 2. The ResNet block had the multi-layers of Conv(3, 3, 256, 256)-IN-ReLu-Conv(3, 3, 256, 256)-IN, where the first and second numbers denoted the 2-D kernel size of a Conv layer, and the third and fourth numbers denoted the channel sizes of the input and output, respectively. The final output of the ResNet block was the summation of the input and output of the multi-layers. Four cascaded ResNet blocks were used in the image translation network.

### B. Dimension Reduction (DR) Network

The $G_{dr}$ consisted of Conv(3, 3, 4, 16)-tanh-Conv(3, 3, 16, 8)-tanh-Conv(3, 3, 8, 3) and the $F_{dr}$ consisted of Conv(1, 1, 3, 8)-tanh-Conv(1, 1, 8, 16)-tanh-Conv(1, 1, 16, 4). In $F_{dr}$, the convolutional filters with the kernel size of 1 induced the constraint that the generated features maintained the original spatial information of input images.

### C. Density Estimation (DE) Model

The $G_{de}$ in the DE model consisted of fully connected layers (FC) as follows: FC(7,8)-tanh-Drop(0.5)-FC(8,**6**)-Softmax, where the first and second numbers in a FC layer meant the input and output sizes of neurons, respectively. Here, Drop(0.5) meant the dropout layer with a retention probability of 0.5 [3]. Six Gaussian mixtures were used for density estimation of given features.

### D. Evolution of the Loss Functions

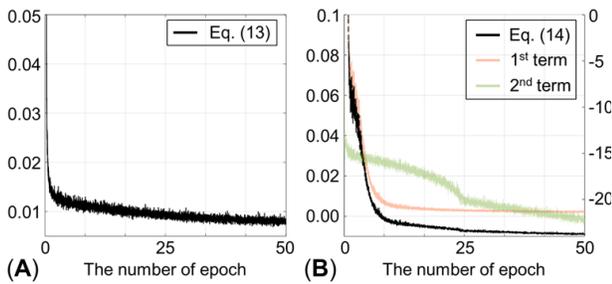

**Fig. S1.** Evolutions of the loss functions in Eq. 13 (A) and Eq. 14 (B), respectively. The ADM network was trained with BraTS dataset for 50 epochs.

## II. EFFECT OF HYPER-PARAMETERS ON THE ADM NETWORK

We performed additional experiments to analyze how hyper-parameters in the ADM network affect performances. Two representative hyper-parameters were selected, which were λ of Eq. 14 and the number of Gaussian mixture models. The AUC values for the BraTS and ISLES test data were measured according to the values of the hyper-parameters (Table S1 and S2). The λ value was used to balance two different losses in Eq. 14. The number of GMM regulated the representation power of the DE network for normal tissues. In case of λ, the AUC had the highest value at 0.0005 for BraTS and 0.005 for ISLES, which was in line with validation data. However, the ADM was not sensitive to λ and worked in a consistent way. That is because the two different losses may not work against each other. For the number of GMM, the ADM method showed the highest performance at 6 for both BraTS and ISLES. But, even for different number of GMM, the ADM network worked stably and was still ranked 1st compared to the competing methods (Table I and II). It seems that the DR network makes the estimated features fit well with the Gaussian model pre-defined by the hyper-parameter.

**Table S1. AUC values for the test data of BraTS and ISLES with respect to λ**

| λ | BraTS | ISLES |
|---|---|---|
| 5e-5 | 0.937 | 0.840 |
| 5e-4 | 0.939 | 0.852 |
| 5e-3 | 0.922 | 0.871 |
| 5e-2 | 0.925 | 0.852 |

**Table S2. AUC values for the test data of BraTS and ISLES with respect to the number of GMM**

| GMM number | BraTS | ISLES |
|---|---|---|
| 2 | 0.935 | 0.830 |
| 4 | 0.931 | 0.860 |
| 6 | 0.939 | 0.871 |
| 8 | 0.936 | 0.846 |

## III. CONFIGURATIONS OF BASELINE MODELS

### A. OC-SVM

OC-SVM adopting the radial basis function kernel was applied to pixel-wise anomaly detection. An input data was a vector containing the pixel intensities of multi-contrast images. That is, the input was a four-dimensional vector for four multi-contrast images. After excluding black-background regions in MR images, 26,443,569 and 3,417,807 vectors were used to train the model for BraTS and ISLES, respectively. The kernel coefficient of the radial basis function was 0.25, which was determined by 1/(the dimension of a feature) [4]. The parameter of $v$ that usually represented the fraction of support vectors should be determined. With the validation data, the value of $v$ was searched in the range of 1e-4 ~ 1e-3 with a step size of 1e-4 for obtaining better performance. The values of $v$ were set to 5e-4 and 6e-4 for BraTS and ISLES, respectively.

### B. GMVAE

For GMVAE, the number of Gaussian models and the weight value for loss functions are important hyper-parameters. For better performance, the two hyper-parameters were empirically determined by performing grid searches in the range of 3 ~ 9 with an increment of 3 for the number of Gaussian models and in the range of 0 ~ 9 with an increment 1 for the weight value. By evaluating AUC values for validation data, the number of Gaussian models was chosen as 3 for BraTS and 6 for ISLES. The weight value was 7 for BraTS and 6 for ISLES. The other experimental conditions followed the implementation details reported in the paper [5].

### C. DSVDD

Similar to OC-SVM, a feature vector containing the pixel intensities of multi-contrast images was used for input data in DSVDD. Since the input data was different from the data used in the original paper [6], we adopted a different kernel size of convolutional layers. However, the number of layers was maintained as three, identical to the original paper. The neural network consisted of Conv(1, 1, 4, 8)-BN-lReLU-Conv(1, 1, 8, 8)-BN-lReLU-Conv(1, 1, 8, 3), where BN meant batch normalization [7] and lReLU meant leaky ReLU [8]. Excepting the kernel size, all experimental conditions followed the implementation details reported in the paper [9].

### D. fanoGAN

We used the same network configurations as the paper [10], except that the spatial resolution was reduced to 64×64 due to a lack of memory storage. For performance evaluation, resultant anomaly score maps were interpolated with a bi-cubic kernel to make 128×128 resolution. Detailed implementations followed the paper [10].

### E. DAG

As with GMVAE, DAG also adopted an autoencoder for feature extraction. The dimension of latent space was set to three through experiments. Six Gaussian mixture models were assumed, identical to the proposed DE module. Other configurations followed the implementation details in the paper [11].

### F. DAG$_c$

DAG was originally developed with FC layers. For fair comparison, the DAG model was modified by using convolutional layers. The DAG$_c$ consisted of Conv(3, 3, 4, 32)-tanh-Conv(3, 3, 32, 16)-tanh-Conv(3, 3, 16, 8)-tanh-Conv(3, 3, 8, 3)-Conv(1, 1, 3, 8)-tanh-Conv(1, 1, 8, 16)-tanh-Conv(1, 1, 16, 32)-tanh-Conv(1, 1, 32, 4). Other conditions were identical to DAG.

## IV. NORMAL IMAGE AUGMENTATION FOR ISLES

We performed additional experiments by merging the normal images of the different datasets and augmenting the training data for segmenting ischemic stroke lesion areas. Two kinds of training datasets were constructed by using three common contrasts in BraTS and ISLES: 1) 828 sets from ISLES and 2) 8618 sets from ISLES and BraTS. The three contrast images of ISLES test data were used for evaluation. Augmentation of the training data generally improved the segmentation performance of ischemic stroke lesion areas, as shown in the AUC and precision-recall curves (Fig. S2(A,B)). Anomaly and normal intensities are distinctly separated in 'ISLES + BraTS', while they are clustered around low intensity in 'ISLES' (Fig. S2(C)).

## V. TWO STRATEGIES TO HANDLE THE DIVERSITY OF PUBLIC MRI DATASETS

The proposed method employed two strategies to handle the contrast diversity of open MRI datasets (BraTS, ISLES). One was to use a median value map as an input of the CtoC network and the other was augmentation with random scaling of image intensity. To validate the effectiveness of the two strategies, we performed an additional experiment for the variants of the ADM with or without the two strategies. The results are shown in Table S3 with AUC metric.

**Table S3. AUC values for BraTS test data by using ADM variants. Note that S1 means the strategy of using a median input value and S2 means the strategy of augmentation with random scaling of image intensity.**

|  | AUC |
| --- | --- |
| ADM wo S1 and S2 | 0.805 |
| ADM wo S2 | 0.918 |
| ADM | 0.939 |

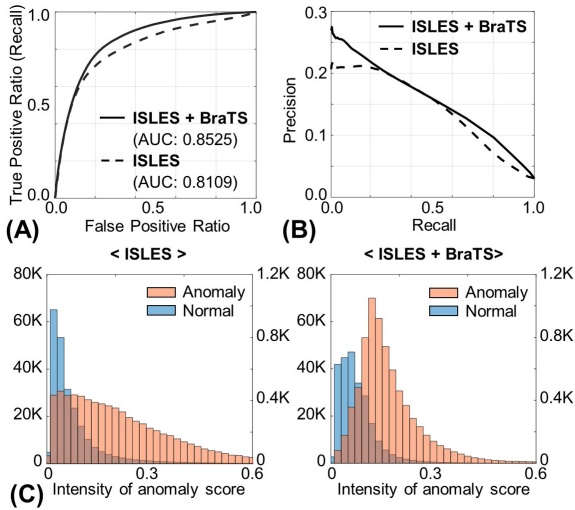

**Fig. S2. Comparison results between 'ISLES' database and 'ISLES + BraTS' database. (A) ROC curves, (B) precision-recall curves and (C) histograms of true normal and anomaly intensities in anomaly score maps. Left and right y axes mean the pixel numbers of normal and anomaly, respectively.**